\documentclass[12pt]{article}
\usepackage[utf8]{inputenc}
\usepackage{amsmath}
\usepackage{amssymb}
\usepackage{graphicx}

\newcommand{\del}{\partial}

\makeatletter

\DeclareTextSymbolDefault{\textquotedbl}{T1}

\numberwithin{equation}{section}

\@ifundefined{date}{}{\date{}}

\usepackage{braket}
\usepackage{bm}
\usepackage{bbm}

\usepackage{epsfig}
\usepackage{pstricks}
\usepackage{color}
\usepackage{tikz}

\usepackage{hyperref}

\newcommand {\nn} {\nonumber}
\newcommand{\bbR}{{\mathbb R}}
\newcommand{\bbC}{{\mathbb C}}

\@ifundefined{showcaptionsetup}{}{%
 \PassOptionsToPackage{caption=false}{subfig}}
\usepackage{subfig}
\makeatother

\begin{document}
\begin{titlepage}
\renewcommand{\thefootnote}{\fnsymbol{footnote}}

\begin{flushright} 
  KEK-TH-2380
\end{flushright} 


\begin{center}
  {\bf \large Backpropagating Hybrid Monte Carlo algorithm\\
    for fast Lefschetz thimble calculations}
\end{center}



\begin{center}
         Genki F{\sc ujisawa}$^{1)}$\footnote
          { E-mail address : fujig@post.kek.jp},
         Jun N{\sc ishimura}$^{1,2)}$\footnote
          { E-mail address : jnishi@post.kek.jp},\\
         Katsuta S{\sc akai}$^{2)}$\footnote
          { E-mail address : sakaika@post.kek.jp} and
         Atis Y{\sc osprakob}$^{1)}$\footnote
          { E-mail address : ayosp@post.kek.jp}


\vspace{1cm}

$^{1)}$\textit{Department of Particle and Nuclear Physics,}\\
\textit{School of High Energy Accelerator Science,}\\
{\it Graduate University for Advanced Studies (SOKENDAI),\\
1-1 Oho, Tsukuba, Ibaraki 305-0801, Japan} 

~

$^{2)}$\textit{KEK Theory Center,
Institute of Particle and Nuclear Studies,}\\
{\it High Energy Accelerator Research Organization,\\
1-1 Oho, Tsukuba, Ibaraki 305-0801, Japan} 
\end{center}

\vspace{0.5cm}

\begin{abstract}
  \noindent
  The Picard-Lefschetz theory has been attracting much attention
  as a tool to evaluate
  a multi-variable integral with a complex weight,
  which appears in various important problems in theoretical physics.
  The idea is to deform   
  the integration contour based on Cauchy's theorem
  using the so-called gradient flow equation.
  In this paper, we propose a fast Hybrid Monte Carlo algorithm
  for evaluating the integral,
  where we ``backpropagate'' the force of the fictitious Hamilton dynamics
  on the deformed contour to that on the original contour,
  thereby reducing the required
  computational cost by a factor of the system size.
Our algorithm can be readily
extended to the case in which one integrates over the flow time
in order to solve not only the sign problem but also 
the ergodicity problem that occurs
when there are more than one thimbles contributing to the integral.
%
This enables, in particular,
efficient identification of
all the dominant saddle points and the associated thimbles.
We test our algorithm
by calculating the real-time evolution of the wave function
  using the path integral formalism.
%
\end{abstract}
\vfill
\end{titlepage}
\vfil\eject


\setcounter{footnote}{0}

\section{Introduction}

Many problems in theoretical physics
are formulated
in terms of a multi-variable integral
with some weight, where the number of variables
or the system size is typically very large.
If the weight is positive semi-definite, we
can regard it as a probability distribution
and perform Monte Carlo calculation based on the idea of importance sampling.
However, it is not straightforward to extend this method
to the case with a complex weight.
For instance, if we use the absolute value of the complex weight
as the probability distribution for sampling and take into account
the phase factor by reweighting, huge cancellation occurs among
the sampled configurations.
For this reason, one needs exponentially large numbers of configurations
as
the system size
increases in order to obtain results
with sufficient precision.
This is the sign problem,
which has been hindering the development of theoretical physics
in various branches.

The situation
regarding the sign problem
has changed drastically in the last decade, however.
There is by now a bunch of new techniques
that have proven to work beautifully in certain classes of problems.
For instance, the tensor renormalization group 
\cite{Levin:2006jai, PhysRevB.86.045139, PhysRevLett.115.180405, Adachi:2019paf, Kadoh:2019kqk}
is applicable if one can reformulate the problem in terms of
a network of tensors, where the indices of tensors are contracted
with those of the adjacent tensors.
This method does not rely on
the
importance sampling,
and hence the sign problem is absent from the outset.

Another promising direction is to complexify the integration variables
keeping the holomorphicity of the weight and the observables.
There are roughly two methods that belong to this category.
One of them is the complex Langevin method 
(CLM) \cite{Klauder:1983sp, Parisi:1984cs, Aarts:2009uq, Aarts:2011ax, Nagata:2015uga, Nagata:2016vkn}, 
which can be viewed
as a naive extension of the ordinary Langevin method
for positive semi-definite weights.
In this method, the expectation values of observables
can be obtained with configurations
generated by solving the complex Langevin equation, which involves
the Gaussian noise term
and the drift term representing the derivative of the weight.
The computational cost is comparable to that of ordinary Monte Carlo methods
for positive semi-definite weights.
The validity of this method requires, however, that the probability distribution
of the drift term and that of the observables should fall off sufficiently fast
at large values, which is not always
satisfied \cite{Nagata:2015uga, Nagata:2016vkn}.
While the method is definitely worth trying for its
simplicity
and its capability for large
system size (See, for instance,
Refs.~\cite{Sexty:2019vqx,Berger:2019odf,Anagnostopoulos:2020xai,Scherzer:2020kiu,Attanasio:2020spv,Ito:2020mys} for semi-realistic calculations.),
one needs to have the luck of satisfying the validity condition
in order to
explore the most interesting parameter regime.

The other method based on complexification of integration variables
uses Cauchy's theorem and
deforms the integration contour in such a way that the
integrand does not oscillate much in phase.
One can then apply standard Monte Carlo methods
to perform the integration along the deformed contour.
The crucial question
is how to find an optimal deformed contour.
A rigorous
answer to this question is provided
by the Picard-Lefschetz theory\footnote{This forms
  the basis of the resurgence theory, which is an attempt to
  extract full non-perturbative information from perturbative 
expansions (See, for instance, Ref.~\cite{Aniceto:2018bis}.).
  It has attracted attention also in the context of quantum cosmology
  since it provides a unique
    prescription to make
    the original oscillatory path integral of Lorentzian quantum gravity
     well defined \cite{Feldbrugge:2017kzv,Jia:2021xeh}.}.
This theory asserts that
an optimal deformed contour
is given by
a certain set of Lefschetz thimbles, which represent the steepest descent
contours in the complexified configuration space originating
from saddle points of the complex 
weight \cite{Witten:2010cx,Cristoforetti:2012su,Cristoforetti:2013wha,Fujii:2013sra}.
This set of thimbles can be obtained
by solving the gradient flow equation,
which involves the gradient of the complex weight,
starting from a
point on the original contour \cite{Alexandru:2015sua}.

In practical applications,
it turns out beneficial to keep finite the ``flow time'',
the amount of the contour deformation by the flow equation,
so that one obtains an integration contour which interpolates
the original contour and the set of Lefschetz 
thimbles \cite{Alexandru:2015sua}.
In particular, this avoids the ergodicity problem
or the multi-modality problem
that arises when there are more than one thimbles 
contributing to the integral,
which are either disconnected or connected at a zero of 
the complex weight \cite{Alexandru:2016ejd}.
Furthermore, in order to solve the sign problem and the ergodicity problem
at the same time,
it has been proposed\footnote{Historically, this idea was born out
  of the previous proposal based on 
tempering \cite{Fukuma:2017fjq,Alexandru:2017oyw,Fukuma:2019uot}, 
  which amounts
  to sampling configurations on many integration contours in parallel,
  swapping the configurations after a fixed interval with certain probability.}
to perform
sampling
on the ``worldvolume'' \cite{Fukuma:2020fez},
which consists of a one-parameter family of the integration contour
obtained 
for various flow time.
Including all these ideas,
we use the word ``the generalized (Lefschetz) thimble 
method'' (GTM) (See Ref.~\cite{Alexandru:2020wrj} for a review.)
to refer to the Monte Carlo method
based on the Picard-Lefschetz theory.

While the GTM does not have the validity issue that exists in the CLM,
its crucial disadvantage concerns the computational cost.
Solving the gradient flow equation is similar in this respect
to solving the complex Langevin equation.
However, one still needs to search for the configurations
on the deformed contour that
make important contribution to the integral.
If one uses the plain Metropolis algorithm to do this,
one cannot move fast in the configuration space
keeping the acceptance rate reasonably high,
which necessarily limits the system size to be small.

In this paper, we propose a fast
Hybrid Monte Carlo (HMC) algorithm that can be used
in the GTM.
In this algorithm, one updates the configuration
by solving a fictitious Hamilton equation of motion
using the gradient of the absolute value of the weight
as the force so that one can move very efficiently
in the configuration space.
In fact, various HMC algorithms were proposed 
in the process of developing the
GTM \cite{Fujii:2013sra,Fukuma:2019uot,Fukuma:2020fez}.
These algorithms solve the Hamilton equation of motion
\emph{on the deformed contour} \cite{Fujii:2013sra,Fukuma:2019uot}
or \emph{on the worldvolume} \cite{Fukuma:2020fez},
which we shall refer to as the deformed manifold collectively in what follows.
This requires complicated procedures of determining a constrained motion 
on the deformed manifold, which is unknown {\it a priori}.
In addition to the force obtained by taking the derivative
of the modulus of the complex weight,
one has the normal force characteristic to constrained systems,
which has to be determined in such a way that
the complexified configuration is
constrained on the deformed manifold.
This is done by Newton's method, which requires
solving the flow equation at each iteration.



Here we consider solving the Hamilton equation of motion
\emph{on the original contour},
which saves us from all the burden associated with treating
the constrained motion.
The main task is transferred to the calculation of the force,
which requires 
differentiating
the modulus
of the complex weight
evaluated on the deformed manifold, now with respect to
the configuration on the original contour\footnote{See
Ref.~\cite{Ulybyshev:2019fte} for an earlier work,
where the force
is calculated by numerical derivative.
}.
At first sight, one might think that this requires
the calculation of the Jacobian matrix associated with the
deformation of the integration contour\footnote{See, for instance,
footnote 30 of Appendix D in the published
version of Ref.~\cite{Fukuma:2020fez}.}.
The crucial point of our proposal is that one can actually
avoid this calculation
by ``backpropagating''
the force
on the deformed contour to that on the original contour,
which reduces the computational cost by
a factor of
the system size.

The only drawback of our HMC algorithm compared with
the existing ones is that the modulus of the Jacobian is not
included in sampling and hence should be
taken into account by reweighting.
Note, however, that the phase of the Jacobian
has to be taken into account by reweighting anyway
in the GTM in general.
The calculation of the Jacobian is time-consuming,
but it can be done off-line
since it is needed only
when one measures the observables
for statistically decorrelated configurations obtained by the HMC algorithm.
If one uses a crude estimator of the Jacobian \cite{Alexandru:2016lsn},
the computational cost of the GTM
becomes
comparable to
ordinary Monte Carlo methods just like the CLM.
%
Thus our algorithm
opens up a new possibility
for
applying the GTM to large systems.

The rest of this paper is organized as follows.
In Section \ref{sec:GTM}, we briefly review the GTM.
In Section \ref{sec:HMC}, we present our HMC algorithm
on the original contour.
In Section \ref{sec:integrate-tau}, we extend our algorithm to the case in which
one integrates over the flow time to avoid the ergodicity problem.
In Section \ref{sec:relationship},
we clarify the relationship to the existing HMC algorithms.
In Section \ref{sec:examples}, we
test
our algorithm
by calculating the real-time evolution of the wave function
using the path integral formalism.
Section \ref{sec:summary} is devoted to a summary and discussions.
In Appendix \ref{sec:bp-jacobian},
we
provide a simple
understanding
for the reduction of computational cost by backpropagation.
In Appendices \ref{sec:details} and \ref{sec:imag-JdagJ-nonzero},
we provide some notes related to Section \ref{sec:relationship}.

\section{Brief review of the GTM}
\label{sec:GTM}

In this section, we briefly review the GTM
for
a general model given by
the partition function
\begin{align}
  Z=\int
  dx \, e^{-S(x)} \ ,
\label{Z}
\end{align}
where the action $S(x)$ is a complex function of 
$x=(x_1,\cdots,x_N) \in \mathbb{R}^N$.
The expectation value of an observable $\mathcal O(x)$ is defined by
\begin{align}
\langle \mathcal O(x)\rangle =
\frac{1}{Z}\int
dx \, \mathcal O(x) \, e^{-S(x)} \ .
\label{O}
\end{align}

In the GTM \cite{Alexandru:2015sua}, 
we deform the integration contour into $\mathbb{C}^N$
by using the so-called holomorphic gradient flow equation
%
%
\begin{align}
  \frac{\partial}{\partial \sigma}z_k(x,\sigma)
= \overline{
  \frac{\partial S(z(x,\sigma))}{\partial z_k}
} \ ,
\label{hge}
\end{align}
which is solved from $\sigma = 0$ to $\sigma=\tau$
with the initial condition $z(x,0)=x\in \mathbb{R}^N$.
The flowed configurations define
an $N$-dimensional real manifold embedded in $\mathbb{C}^N$,
which we denote as $M_\tau=\{z(x,\tau)|x\in\mathbb{R}^N\}$.
According to Cauchy's theorem,
the integration contour can be deformed continuously
from $\mathbb{R}^N$ to $M_\tau\subset \mathbb{C}^N$
without changing the partition function.
Then, by noting that 
$z = z(x,\tau)$
defines a one-to-one map
from $x\in \mathbb{R}^N$ to $z\in M_\tau$,
one can rewrite the partition function as
\begin{align}
  Z
=\int_{M_\tau} dz \, e^{-S(z)}
=\int
dx \, e^{-S(z(x,\tau))} \, \mathrm{det}J(x,\tau) \ , 
\label{gltmZ}
\end{align}
and similarly the observable \eqref{O} as
\begin{align}
  \langle \mathcal O(x)\rangle
  =\frac{1}{Z} \int dx \, \mathcal O(z(x,\tau))
e^{-S(z(x,\tau))} \, \mathrm{det}J(x,\tau)  \ ,
\label{def_O-Z}
\end{align}
where
the Jacobian matrix $J(x,\tau)$ corresponding to the map $z(x,\tau)$
is defined by
\begin{align}
J_{kl}(x,\sigma) \equiv \frac{\partial z_k(x,\sigma)}{\partial x_l} \ .
\label{J-def}
\end{align}
Taking the derivative with respect to $x_l$ on each side of \eqref{hge},
one obtains
the differential equation for $J(x,\sigma)$ as
\begin{align}
\frac{\partial}{\partial \sigma}J_{kl}(x,\sigma)
&= \overline{H_{km}(z(x,\sigma))} \,
\overline{J_{ml}(x,\sigma)} \ ,
\label{hgeJ}
\end{align}
which can be solved
from $\sigma = 0$ to $\sigma=\tau$
with the initial condition $J_{kl}(x,0)=\delta_{kl}$
to obtain the Jacobian matrix $J(x,\tau)$.
Here we have defined the Hessian of the action as
\begin{align}
    H_{ij}(z)  &= \frac{\partial^2 S(z)}{\partial z_i\partial z_j} \ ,
\label{def-H}
\end{align}
which plays an important role in this paper.
Note that the Hessian is a sparse matrix with only O($N$) nonzero elements
if the action is \emph{local}, meaning that
only terms with local couplings among $z_i$'s exist
as in the case of local field theories.

The virtue of using the holomorphic gradient flow
\eqref{hge} in deforming the integration contour
can be understood
by taking the derivative of the action
with respect to the flow time as
\begin{align}
  \frac{\del}{\del \sigma} S(z(x,\sigma))
  &= \frac{\del z_k (z(x,\sigma))}{\del \sigma} \,
  \frac{\del S (z(x,\sigma))}{\del z_k}
  =  \overline{\frac{\del S (z(x,\sigma))}{\del z_k}} \,
  \frac{\del S (z(x,\sigma))}{\del z_k}
    \ge 0 \ ,
\label{derivative-S}
\end{align}
which implies that
the real part of the action $S$ grows monotonically along the flow
while keeping the imaginary part constant.
Thus, in the $\sigma \rightarrow \infty $ limit,
${\rm Re} S (z(x,\sigma))$
diverges except for 
such points $x=x^\star$ that satisfy
$\lim_{\sigma \rightarrow \infty}
\frac{\del}{\del z_k} S(z(x^\star ,\sigma)) = 0$,
which are
the points that flow into
some fixed points $z=z^\star$
defined by $\frac{\del S(z^\star)}{\del z_k}  = 0$.
The deformed contour $M_\tau$
at $\tau = \infty$
consists of
a set of Lefschetz thimbles associated with these fixed points,
which correspond to the infinitesimal vicinities of the
points $x^\star$ on the original contour.
Since $\mathrm{Im}S(z(x))$ is constant 
on each thimble, the sign problem is maximally reduced.
The original proposal was to perform
Monte Carlo integration
on these
thimbles \cite{Cristoforetti:2012su,Cristoforetti:2013wha,Fujii:2013sra}.

However, 
when there are more than one thimbles,
the transition between thimbles does not occur 
during the Monte Carlo simulation,
which leads to the violation of 
ergodicity.
The GTM \cite{Alexandru:2015sua}
avoids this problem
by choosing a sufficiently small flow time
allowing the $\mathrm{Im}S(z(x))$ to fluctuate to some extent,
which works as far as the system size is not so large.
For a larger system size, one can
integrate over the flow time in order to avoid
the sign problem and the ergodicity problem
at the same time \cite{Fukuma:2020fez}.
%


\section{HMC algorithm with backpropagation}
\label{sec:HMC}

In this section, we discuss our HMC algorithm on the original contour.
The crucial point is
to calculate
the force in the
fictitious Hamilton dynamics
by using the idea of backpropagation,
which reduces the computational cost by a factor of O($N$).
Here we consider the case of fixed $\tau$
and defer the discussion in the case of integrating $\tau$
to Section \ref{sec:integrate-tau}.


\subsection{HMC algorithm on the original contour}
\label{sec:HMC-GTM}

In order to
simulate
the partition function \eqref{gltmZ},
let us consider the partition function
\begin{align}
Z_\tau
=\int_{\mathbb{R}^N} dx \, e^{-{\rm Re} S(z(x,\tau))} \ ,
\label{def_Z0}
\end{align}
where the imaginary part of the action $S$ as well as the Jacobian is omitted.
The expectation value \eqref{def_O-Z}
can be obtained
by the standard reweighting formula as
\begin{align}
\langle \mathcal O(x)\rangle
=\frac{\langle \mathcal O(z(x,\tau))
  e^{-i\mathrm{Im}S(z(x,\tau))} \mathrm{det}J(x,\tau)
\rangle_\tau }
{\langle e^{-i\mathrm{Im}S(z(x,\tau))}\mathrm{det}J(x,\tau) \rangle_\tau  }
\ ,
\label{gltm_ob}
\end{align}
where $\langle\cdots\rangle_\tau$ represents 
the expectation value
with respect to the partition function \eqref{def_Z0}.

The most naive way to simulate the
partition function \eqref{def_Z0}
is the Metropolis algorithm.
Starting from some configuration $x \in \mathbb{R}^N$,
one constructs a trial configuration $x' =  x + \delta x \in \mathbb{R}^N$
with certain probability distribution for $\delta x$.
Then one calculates the change of the action
\begin{align}
  \delta S =   \mathrm{Re} S(z(x',\tau)) -  \mathrm{Re} S(z(x,\tau)) \ ,
\label{def-deltaS}
\end{align}
and accepts the trial configuration with the probability $\min (1, e^{-\delta S})$.
Both the numerator and the denominator in \eqref{gltm_ob}
can be obtained by taking an average over the configurations
generated by the above algorithm.
While this algorithm is very easy to implement, it is not efficient
since the update by $\delta x$ has to be
local
in order to keep
the acceptance rate reasonably high.

The basic idea of the HMC algorithm \cite{Duane:1987de}
is to enable a global update $\delta x$
in the Metropolis algorithm
by
using the information of the action in its proposal.
For that, we introduce auxiliary variables
$p_i$ corresponding to $x_i$ and consider the partition function
\begin{align}
  \tilde{Z}_\tau
  \label{Z-HMC}
  &=\int
  dx \, dp  \, e^{-H}  \ , \\
H &= \frac{1}{2} (p_i)^2 + {\rm Re} S(z(x,\tau)) \ ,
\label{def_Ham}
\end{align}
which is equivalent to \eqref{def_Z0}.
One can update $p_i$ by just generating Gaussian random numbers.
After that,
using the initial condition for $x(0)$ and $p(0)$
set by the current configuration,
one
solves the fictitious Hamilton equation
\begin{align}
  \frac{dx_i(s)}{ds} &=
  p_i(s) \ , \nn
  \\
  \frac{dp_i(s)}{ds} &=
  F_i(s) \ ,
\label{Ham-eq}
\end{align}
where the force $F_i(s)$ is defined by
\begin{align}
  F_i(s) &= - \left. \frac{\del {\rm Re}S(z(x,\tau))}{\del x_i} \right|_{x=x(s)}
\label{def-F}
\end{align}
for a fixed time $s_{\rm f}$
to obtain
$x(s_{\rm f})$ and $p(s_{\rm f})$,
which provide the trial configuration
subject to the Metropolis accept/reject procedure with the probability
$\min (1 , e^{- \delta H})$, where
\begin{align}
\delta H & = H(x(s_{\rm f}),p(s_{\rm f})) - H(x(0),p(0))  \ .
\label{dela-H}
\end{align}
If one can solve \eqref{Ham-eq} exactly, one obtains $\delta H=0$ due to
the Hamiltonian conservation, which implies that the acceptance rate
is 100\%.
In practice, one has to discretize the Hamilton equation,
which makes $\delta H \neq 0$.
The stepsize $\Delta s$ has to be chosen small enough to
keep
the acceptance rate
reasonably high\footnote{More precisely, one can
optimize $\Delta s$ by maximizing the
product of $\Delta s$ and the acceptance rate, which represents
the effective speed.  
The amount of ``time'' $s_{\rm f}$ for which one solves the Hamilton equation
is another tunable parameter of the algorithm, which can be optimized
by minimizing the autocorrelation time in units of the
step of $\Delta s$ one makes in solving the Hamilton equation.}.
Note that the detailed balance is satisfied
even at finite $\Delta s$ thanks to the Metropolis procedure,
and hence there is no systematic error due to the discretization
of the Hamilton equation.
See Section \ref{sec:discretizing-HMC} for more detail.

\subsection{Calculating the force by backpropagation}
\label{sec:force-backpropagation}

Thus the question boils down to the calculation of
\eqref{def-F}.
Let us first note that 
\eqref{def-F} can be rewritten as
\begin{align}
  F_j(s) &=  2 \, {\rm Re}
  \left\{ f_i  J_{ij}(x(s),\tau) \right\} \ ,
\label{def-F2}
\end{align}
where $J_{ij}(x,\tau)$ is the Jacobian matrix defined by
\eqref{J-def} and $f_i(s)$ is the gradient of the action on the deformed contour
\begin{align}
f_i &= - \left. \frac{\del {\rm Re} S(z)}{\del z_i} \right|_{z=z(x(s),\tau)}  
\label{def-f-deformed}
\end{align}
at
$z(x(s),\tau)$
corresponding to the point $x(s)$ on the original contour.
The calculation of the Jacobian matrix, however,
requires us to solve the differential equation \eqref{hgeJ},
which involves O($N^2$) or O($N^3$) arithmetic operations
depending on whether the action $S(z)$ is local or non-local
in the sense explained below \eqref{def-H}.
Alternatively,
one might think of evaluating
the right-hand side of \eqref{def-F} by numerical
derivative as in Ref.~\cite{Ulybyshev:2019fte}.
The required cost, however,
is comparable to calculating the Jacobian,
and what is worse, there is some systematic error associated with
the numerical derivative.

Suppose we were to calculate $J_{ij}(x,\tau)\, v_j$
with $v_j \in \bbR$,
which
appears many times in the HMC algorithm on the deformed
manifold \cite{Fukuma:2019uot,Fukuma:2020fez}.
In this case,
one can avoid the calculation of
$J_{ij}(x,\tau)$ itself by noticing that
$v_i(\sigma) \equiv J_{ij}(x,\sigma) \, v_j$ satisfies the flow equation
\begin{align}
\frac{\partial}{\partial \sigma}v_{i}(\sigma)
=\overline{H_{ij}(z(x,\sigma))} \, 
\overline{v_j(\sigma)} \ .
\label{flow-v}
\end{align}
Solving this from $\sigma = 0$ to $\sigma=\tau$
with the initial condition $v_i(0)=v_i$
requires O($N$) or O($N^2$)
arithmetic operations
depending on whether the action $S(z)$ is local or non-local.
Thus one can save a factor of O($N$).

Coming back to our problem,
we can actually save a factor of O($N$) as well
but in a slightly different way.
Let us note that the problem of calculating \eqref{def-F}
is mathematically equivalent to the calculation in machine learning,
where ${\rm Re} S(z(x,\tau))$ represents the loss function
and $z_i(x,\tau)$ represents the parameters of the network at a layer labeled
with $\tau$.
The quantity \eqref{def-F}
we have to calculate
is the derivative of the loss function
with respect to $x_i = z_i(x,0)$,
which represent the parameters at the first layer.
The crucial idea here is the use of backpropagation.\footnote{The use of
  backpropagation in calculating the force in HMC algorithms is actually quite common
  in lattice gauge theory when the action $S$ is written in terms of smeared
  field variables, which are obtained by solving some gradient flow equation
  starting from the original field variables.
  We would like to thank Akio Tomiya for bringing our
  attention to their work \cite{Tomiya:2021ywc} in this regard.}
Let us define
the ``force'' at flow time $\sigma$ by
\begin{align}
f_i(\sigma) &= -  \frac{\del {\rm Re} S(z(x,\tau))}{\del z_i(x,\sigma)}  \ .
\label{def-f-sigma}
\end{align}
Applying the chain rule, we obtain 
\begin{align}
  f_j (\sigma -\varepsilon)
  &=    f_i (\sigma) \frac{\del z_i(x,\sigma)}{\del z_j(x,\sigma-\varepsilon)}
  +  \overline{f_i (\sigma)}
   \frac{\del \overline{z_i(x,\sigma)}}{\del z_j(x,\sigma-\varepsilon)}  \\
   &\sim   f_j (\sigma) +  \varepsilon \,
   \overline{f_i (\sigma)} \, H_{ij}(\sigma) \ .
   \label{force-backprop-flow-complex0}        
\end{align}
This implies that $f_i(\sigma)$
obeys the differential equation
\begin{align}
  \frac{d}{d \sigma}
  f_j (\sigma)
  &= -   \overline{f_i(\sigma)} \, 
  H_{ij}(\sigma) \ ,
\label{force-backprop-flow-complex}        
\end{align}
which is analogous to \eqref{flow-v}.
By solving this differential equation \emph{backwards in} $\sigma$
from $\sigma=\tau$ to $\sigma=0$
with the initial condition $f_i (\tau) = f_i$,
one obtains the force at the first layer
\begin{align}
  f_i(0)
  &=
  -  \frac{\del {\rm Re} S(z(x,\tau))}{\del z_i(x,0)}  \ .
\label{force-final0}
\end{align}
Using this, the desired force \eqref{def-F} is obtained as
\begin{align}
  F_i
  &=
  -  \frac{\del {\rm Re} S(z(x,\tau))}{\del z_i(x,0)} 
  -  \frac{\del {\rm Re} S(z(x,\tau))}{\del \bar{z}_i(x,0)}
  =  2 \,  {\rm Re} f_i(0) \ .
\label{force-final}
\end{align}
Thus the computational cost is reduced by a factor of O($N$).

\subsection{Discretizing the flow equation}
\label{sec:discretizing-flow-time}

In actual calculations, one has to discretize
the holomorphic gradient flow equation \eqref{hge}.
In this section, we explain how our idea for calculating the force
in the fictitious Hamilton dynamics can be extended to this case.
In particular, we
show that
this can be done
in such a way that the discretization 
causes neither systematic errors
nor any decrease in the acceptance rate,
%
which is not the case for
the existing HMC
algorithms \cite{Fukuma:2019uot,Fukuma:2020fez} as we discuss
in Section \ref{sec:relationship}.
%


Let us
define
$\sigma_n = n \, \varepsilon$ ($n=0, 1 , \cdots , N_{\tau}$)
with $\tau = N_{\tau} \, \varepsilon$
and consider the holomorphic gradient flow \eqref{hge} discretized as 
\begin{align}
z_k(x,\sigma_{n+1})
= z_k(x,\sigma_n)  + 
\varepsilon \,
\overline{
  \frac{\partial S(z(x,\sigma_n))}{\partial z_k}
} 
\label{hge-discretized}
\end{align}
with the initial condition $z(x,0)=x\in \mathbb{R}^N$.
Taking the derivative with respect to $x_l$ on each side
of \eqref{hge-discretized},
one obtains
\begin{align}
  J_{kl}(x,\sigma_{n+1})
  = J_{kl}(x,\sigma_n) + 
  \varepsilon \, 
    \overline{H_{km}(z(x,\sigma_n))}
  \, 
\overline{J_{ml}(x,\sigma_n)} \ .
\label{hgeJ-discretized}
\end{align}
By solving this difference equation
with the initial condition $J_{kl}(x,0)=\delta_{kl}$,
one obtains \emph{exactly}
the Jacobian matrix \eqref{J-def} defined for the solution 
to the discretized flow equation \eqref{hge-discretized}.
This means that while finite $\varepsilon$ affects
the deformed contour, thanks to Cauchy's theorem, it does
not affect the equations \eqref{gltmZ} and \eqref{def_O-Z},
and hence
does not cause any systematic error.


Let us then discuss how the calculation of the force
\eqref{def-F}
should be extended to the case of the discretized flow equation.
If one uses \eqref{def-F2},
the Jacobian matrix has to be obtained
by solving \eqref{hgeJ-discretized}, which is time consuming.
Instead, we define the ``force'' at flow time $\sigma_n$ by
\begin{align}
f_i(\sigma_n) &= -  \frac{\del {\rm Re} S(z(x,\tau))}{\del z_i(x,\sigma_n)}  \ .
\label{def-f-sigma-discrete}
\end{align}
Applying the chain rule, we obtain
\begin{align}
  f_j (\sigma_{n-1})
  &=    f_i (\sigma_n) \frac{\del z_i(x,\sigma_n)}{\del z_j(x,\sigma_{n-1})}
  +  \overline{f_i (\sigma_{n})}
   \frac{\del \overline{z_i(x,\sigma_n)}}{\del z_j(x,\sigma_{n-1})}  \\
   &=  f_j (\sigma_n) +  \varepsilon \,  \overline{f_i (\sigma_n)} H_{ij}(\sigma_{n-1}) \ .
   \label{force-backprop-flow-complex0-discrete}        
\end{align}
By solving this difference equation \emph{backwards in} $\sigma$
from $\sigma=\tau$ to $\sigma=0$
with the initial condition $f_i (\tau) = f_i$,
one obtains the force at the first layer \eqref{force-final0}.
Using this, the desired force is obtained as \eqref{force-final}.
The computational cost is reduced by a factor of O($N$)
similarly to the case of the continuous flow equation. 
In Appendix \ref{sec:bp-jacobian}, we provide a simple
understanding for this huge reduction of computational cost.

\section{Integrating over the flow time $\tau$}
\label{sec:integrate-tau}

In order to solve the sign problem, the flow time $\tau$ has
to be sufficiently large.
However, when one has more than one Lefschetz thimbles
at $\tau = \infty$, Monte Carlo simulations
based on the HMC algorithm suffer from the ergodicity problem.
This occurs because the regions that contribute to the integral
are either disconnected or separated by singular points, at which
the real part of the action diverges.
In order to solve the sign problem without suffering from this
ergodicity problem, it is useful to
integrate over the flow time.
This is the idea of the ``worldvolume approach'' \cite{Fukuma:2020fez},
which was put forward 
in the context of the HMC algorithm on the deformed contour.
Here we adapt this idea to our HMC algorithm on the original contour,
which not only turns out to be much simpler but also solves a possible problem
in the original proposal.



\subsection{The basic idea}
\label{sec:integrate-tau-basic}

The key observation here is that
in the reweighting formula \eqref{gltm_ob},
the ratio is $\tau$-independent
although
both the numerator and the denominator on the right-hand side
depend on $\tau$.
This implies that one can actually integrate over $\tau$
and
consider
\begin{align}
Z_W
=\int d\tau \, e^{-W(\tau)} \int dx \, e^{-{\rm Re} S(z(x,\tau))} \ ,
\label{def_Z_W}
\end{align}
where $W(\tau)$ is a real function of $\tau$ that can be chosen
so that the distribution of $\tau$ becomes almost uniform
within a certain range \cite{Fukuma:2020fez}.
The expectation value of an observable can be calculated as
\begin{align}
\langle \mathcal O(x)\rangle
=\frac{\langle \mathcal O(z(x,\tau))
  e^{-i\mathrm{Im}S(z(x,\tau))} \mathrm{det}J(x,\tau)
 e^{- \widetilde{W}(\tau)}
\rangle_W }
{\langle e^{-i\mathrm{Im}S(z(x,\tau))}\mathrm{det}J(x,\tau) 
e^{- \widetilde{W}(\tau)}
  \rangle_W  }
\ ,
\label{gltm_ob-tau}
\end{align}
where $\langle\cdots\rangle_W$ represents 
the expectation value
with respect to the partition function \eqref{def_Z_W}.
We have introduced an arbitrary function $\widetilde{W}(\tau)$,
which can be chosen to minimize
the statistical errors \cite{Fukuma:2021aoo}.

\subsection{Extending our HMC algorithm}
\label{sec:integrate-tau-HMC}

Let us extend our HMC algorithm to the model \eqref{def_Z_W}.
As we did in the previous section,
we introduce auxiliary variables
$(p_i,p_\tau)$ corresponding to $(x_i,\tau)$ and
consider the partition function
\begin{align}
  \tilde{Z}_W
  \label{Z-HMC-W}
  &=\int d\tau \, dp_\tau \, dx  \,  dp  \, e^{-H}  \ , \\
  H &= \frac{1}{2} (p_i)^2 + \frac{1}{2} (p_\tau)^2
  + {\rm Re} S(z(x,\tau))  +  W(\tau) \ .
\label{def_Ham-W}
\end{align}

The Hamilton equation for $x_i(s)$ and $p_i(s)$ are given by \eqref{Ham-eq}
as before, but here we also have the equations for $\tau(s)$ and $p_\tau(s)$
given by
\begin{align}
  \label{Ham-eq-tau0}
  \frac{d\tau(s)}{ds} &=
  p_\tau(s) \ , \\
  \frac{dp_\tau(s)}{ds} &=
   -   \left. \frac{d W(\tau)}{d \tau} \right|_{\tau = \tau(s)} 
  + F_\tau(s) \ ,
\label{Ham-eq-tau}
\end{align}
where the force $F_\tau(s)$ for $\tau$ is defined by
\begin{align}
  F_\tau(s) &= 
- \left. \frac{\del {\rm Re}S(z(x,\tau))}{\del \tau} \right|_{x=x(s),\tau=\tau(s)} \ .
  \label{def-F-tau0}
\end{align}
Using the flow equation \eqref{hge}, one can rewrite it as
\begin{align}
    F_\tau(s) &=
-  \left| \left. \frac{\del S(z)}{\del z_i}\right|_{z=z(x(s),\tau(s))} \right|^2 \ .
\label{def-F-tau}
\end{align}
%


\subsection{Discretizing the flow equation keeping $\tau$ continuous}
\label{sec:tau-discretized}

In actual calculations, we have to discretize the flow equation
by $\tau = N_\tau \, \varepsilon$.
The integration over $\tau$ should then be replaced by the integration
over the flow stepsize $\varepsilon$,
and the Hamilton equation
\eqref{Ham-eq-tau0}, \eqref{Ham-eq-tau} should be regarded
as describing the fictitious time evolution of the flow stepsize $\varepsilon$.

The calculation of the force \eqref{def-F-tau0} in the $\tau$-direction
requires some care.
Note that \eqref{def-F-tau} is obtained
by using the continuum version of the flow equation \eqref{hge}.
Therefore, if one uses \eqref{def-F-tau} naively, the Hamiltonian conservation
in the HMC algorithm is violated by the discretization of the flow equation.
Instead, we have to calculate \eqref{def-F-tau0} as
\begin{align}
  F_\tau(s)
  &= - \frac{1}{N_\tau}
  \left. \frac{\del {\rm Re}S(z(x, N_\tau \varepsilon))}{\del \varepsilon}
  \right|_{x=x(s),\tau=\tau(s)} 
  \label{def-F-tau0-discretized}
  \\
  &= - \frac{1}{N_\tau}
  {\rm Re} \left\{ 
  \left.
  \frac{\del z_k(x, N_\tau \varepsilon)}{\del \varepsilon}
  \right|_{x=x(s), \, \varepsilon=\tau(s)/N_\tau}
  \left.
  \frac{\del S(z(x, \tau))}{\del z_k}
  \right|_{x=x(s),\tau=\tau(s)}   \right\} 
   \ ,
  \label{def-F-tau0-discretized2}
\end{align}
where $z_k(x, N_\tau \varepsilon)$ is obtained by solving
the difference equation \eqref{hge-discretized}.
Taking the $\varepsilon$-derivative on each side of \eqref{hge-discretized},
where $\sigma_n(\varepsilon)=n \, \varepsilon$ is treated now
as a function of $\varepsilon$, we obtain
\begin{align}
\frac{\del z_k(x, \sigma_{n+1}) }{\del \varepsilon} 
= \frac{\del z_k(x, \sigma_n )}{\del \varepsilon} 
+  \varepsilon
\, 
\overline{
  H_{kj}( \sigma_n ) }
\, 
\overline{
  \frac{\partial z_j(x, \sigma_n)}{\partial \varepsilon}
}
+ 
\overline{
  \frac{\partial S(z(x, \sigma_n))}{\partial z_k}
}
\ .
\label{hge-discretized-3}
\end{align}
Solving this difference equation with the initial condition
\begin{align}
\frac{\partial z_j(x,0)}{\partial \varepsilon} = 0 \ ,
\end{align}
one can obtain the first factor
in \eqref{def-F-tau0-discretized2}.
The computational cost for this procedure
is comparable to that for calculating the force in the $x$-direction.

Note that this is an exact calculation of the force
for finite $\varepsilon$, and hence there is
no systematic error here, either.
This is a significant advantage of our algorithm
over
the existing ones, in which the discretization
of the flow equation causes some systematic error
as we discuss below.

\section{Relationship to the existing HMC algorithms}
\label{sec:relationship}

As we mentioned in the Introduction, there are HMC algorithms
proposed for the GTM in the past.
The difference from our HMC algorithm is that
the existing ones \cite{Fukuma:2019uot,Fukuma:2020fez}
deal with the Hamilton dynamics of
the point $z(x,\tau)$ after the flow,
whereas
we
deal with
the Hamilton dynamics of the point $(x,\tau)$ before the flow.
Since the two points 
are in one-to-one correspondence,
we should be able to compare the Hamilton dynamics.
In this section, we clarify the relationship between the two types
of algorithms from this point of view.
Some basic properties of the deformed manifold
necessary to understand this section
are recapitulated in Appendix \ref{sec:details}
for the readers' convenience.

Here we will discuss the case of integrating $\tau$.
For that, in this Section alone, we introduce the notation
$x_\mu$ and $p_\mu$
($\mu = 0 , 1 , \cdots , N$)
for the dynamical variables and the corresponding momenta,
where we have defined
$x_0 \equiv \tau$
and $p_0 \equiv p_\tau$.
Accordingly, we will also use $z_i(x)$ instead of $z_i(x,\tau)$
for the flowed configuration.
The flow equation can be assumed to be either discretized or continuous,
which does not matter in our discussion below.
%
The above notations enable us
to obtain the results for the case of a fixed flow time $\tau$
by simply replacing the index $\mu= 0 , \cdots  , N$ 
by $i = 1 , \cdots , N$.

The Hamilton dynamics
of the point $z_i(x)$
is described by the Hamiltonian
\begin{align}
  H = \bar{\pi}_i \pi_i + V(z,\bar{z}) \ ,
  \label{H-deformed-contour}
\end{align}
where $\pi_i$ is the momentum on the deformed manifold.
Therefore, the Hamilton equation
on the deformed manifold is given by
\begin{align}
  \frac{dz_i}{ds} &= \pi_i  \ , \label{z-evolve-deformed}
\\
  \frac{d\pi_i}{ds} &=
  - \frac{\del V(z,\bar{z})}{\del \bar{z}_i} + {\cal N}_i \ ,
\label{Ham-eq-deformed}
\end{align}
where ${\cal N}_i$ is
the normal force, which is 
perpendicular to the tangent space of the deformed manifold,
and hence does not affect the Hamiltonian conservation.
The normal force ${\cal N}_i$ is determined in such a way that
the momentum $\pi_i(s)$
resides in the tangent space of the deformed manifold
at the point $z_i(s)$, which is constrained to be on the deformed manifold.

In fact, the Hamilton dynamics
on the deformed manifold
given above
is \emph{not} equivalent to
the Hamilton dynamics on the original contour that we have considered.
Instead, it is described by
the Hamiltonian on the original contour
that takes the form
\begin{align}
H = \frac{1}{2} \, p_\mu K_{\mu\nu}(x) \, p_\nu + V(z(x),\bar{z}(x)) \ ,
  \label{H-equiv}
\end{align}
which 
involves a nontrivial kernel $K_{\mu\nu}(x)=K_{\nu\mu}(x)$
in the kinetic term to be specified later.
The Hamilton equation therefore takes the form
\begin{align}
  \frac{dx_\mu}{ds} &=
  K_{\mu \nu}(x) \, p_\nu \ ,   \\
  \frac{dp_\mu}{ds} &=
  - \frac{\del V(z(x),\bar{z}(x))}{\del x_\mu} -
  \frac{1}{2} \, p_\nu \frac{\del K_{\nu\lambda}(x)}{\del x_\mu} p_\lambda \ .
\label{Ham-eq-equiv}
\end{align} 
The $s$-evolution of the flowed configuration $z_i(x(s))$ is then given by
\begin{align}
  \frac{dz_i}{ds} &=
    \frac{\del z_i}{\del x_\mu} \,  \frac{dx_\mu}{ds}
=  \frac{\del z_i}{\del x_\mu} K_{\mu \nu}(x) \, p_\nu \ .
\label{z-evolve-equiv}
\end{align}
Comparing this with \eqref{z-evolve-deformed},
we obtain the relationship
\begin{align}
\pi_i = \frac{\del z_i}{\del x_\mu} K_{\mu \nu}(x) \, p_\nu \ ,
\label{pi-identify}
\end{align}
which shows that the momentum $\pi_i$ on the deformed manifold
resides in the tangent space
as it should.
Plugging \eqref{pi-identify} and $z_i=z_i(x)$ into
\eqref{H-deformed-contour},
one obtains
\begin{align}
  H &= p_\mu \, K_{\mu \lambda}(x) \, g_{\lambda \rho}(x) \,
  K_{\rho \nu}(x) \, p_\nu + V(z(x),\bar{z}(x))
  \ ,
\label{Ham-deformed-mfd-2}
\end{align}
where we have defined
\begin{align}
g_{\lambda \rho}(x) &= {\rm Re} \left(
  \overline{\frac{\del z_i}{\del x_\lambda}} \, 
  \frac{\del z_i}{\del x_\rho}  \right) \ .
  \label{def-g}
\end{align}
Since \eqref{Ham-deformed-mfd-2}
should be identified with \eqref{H-equiv},
we obtain an identity
$\frac{1}{2} \, K(x) = K(x) \,  g(x) \,  K(x)$,
which implies 
\begin{align}
K(x)=\frac{1}{2}\, g^{-1}(x)  \ .
  \label{K-identity}
\end{align}

Next, let us consider
the $s$-evolution of the momentum $\pi_i(x)$
on the deformed manifold.
From \eqref{pi-identify}, we find
\begin{align}
  \frac{\del \pi_i}{\del s}
  =&
\frac{\del z_i}{\del x_\mu} K_{\mu \nu}(x) \, \frac{\del p_\nu}{\del s} 
+  \frac{\del x_\lambda}{\del s} \frac{\del}{\del x_\lambda} 
\left( \frac{\del z_i}{\del x_\mu} K_{\mu \nu}(x)\right) \, p_\nu
= 
   {\cal F}_i
   + {\cal N}_i '  \ ,
   \label{pi-s-evolve}
\end{align}
where we have defined
\begin{align}
  {\cal F}_i
  &=
-   \frac{\del z_i}{\del x_\mu} K_{\mu \nu}(x) \,
\frac{\del V(z(x),\bar{z}(x))}{\del x_\nu}  \ ,  \\
  {\cal N}_i '
  &=
- \frac{1}{2} \, \frac{\del z_i}{\del x_\mu} K_{\mu \nu}(x) \,
p_\lambda \frac{\del K_{\lambda\rho}(x)}{\del x_\nu} p_\rho
+   K_{\nu \lambda}(x) \, p_\lambda
\frac{\del}{\del x_\nu} 
\left( \frac{\del z_i}{\del x_\mu} K_{\mu \rho}(x)\right) \, p_\rho   \ .
\label{def-FN}
\end{align}
The first term ${\cal F}_i$ 
in \eqref{pi-s-evolve} 
is given by
\begin{align}
  {\cal F}_i
&=
- \frac{\del z_i}{\del x_\mu} K_{\mu \nu}(x) \,
\left( \frac{\del z_j}{\del x_\nu}
\frac{\del V(z,\bar{z})}{\del z_j}
+  \frac{\del \bar{z}_j}{\del x_\nu}
\frac{\del V(z,\bar{z})}{\del \bar{z}_j} \right) \\
&=
-  \frac{\del z_i}{\del x_\mu} \{ g^{-1}(x)\}_{\mu \nu} \,
{\rm Re} \left( \overline{\frac{\del z_j}{\del x_\nu}} \,
\frac{\del V(z,\bar{z})}{\del \bar{z}_j} \right)  \ .
\label{z-force-2}
\end{align}
This is nothing but the gradient force
$- \frac{\del V(z,\bar{z})}{\del \bar{z}_i} $
projected onto the tangent space.
The second term ${\cal N}_i '$
in \eqref{pi-s-evolve} can be interpreted as
the normal force that
is needed to realize the constrained motion.
This can be checked
by confirming that its projection onto the
tangent space vanishes identically; namely,
\begin{align}
  {\rm Re} \left( \overline{\frac{\del z_i}{\del x_\alpha}} \, {\cal N}_i '
 \right) = 0  \quad \quad \mbox{for all $\alpha$} \ .
  \label{project-N}
\end{align}
See Appendix \ref{sec:details} for the details.


The appearance of the kernel $K=\frac{1}{2}\, g^{-1}$
in \eqref{H-equiv} 
can be understood also from the viewpoint
of what we are simulating.
The Hamilton dynamics 
we consider in Section \ref{sec:integrate-tau-HMC}
is intended to reproduce the partition function \eqref{def_Z_W}.
This can be seen by integrating out the momentum variables 
in \eqref{Z-HMC-W}, which yields \eqref{def_Z_W}.
If we have a nontrivial kernel $K=\frac{1}{2}\, g^{-1}$
as in \eqref{H-equiv}, 
the integration over the momentum $p_\mu$ in \eqref{H-equiv}
yields an extra factor $\sqrt{\det g(x)}$
in the partition function.
This is nothing but the volume element of the deformed manifold
since $g(x)$ defined by \eqref{def-g}
represents the induced metric on the deformed manifold,
which is embedded in
$\bbC ^{N}$.
Thus our result \eqref{H-equiv}
is consistent with the fact that the HMC algorithm
using the Hamilton dynamics on the deformed manifold
includes the volume element
of the deformed manifold
in the partition function \cite{Fukuma:2019uot,Fukuma:2020fez}.

In the case of fixed $\tau$, the induced metric 
\eqref{def-g} is given by $g_{kl} = {\rm Re} (J^\dag J)_{kl}$
for the discretized flow equation.
In the continuum limit $\varepsilon \rightarrow 0$ of the
flow equation,
we have ${\rm Im}(J^\dag J) = 0$,
which implies that $\sqrt{\det g(x)}= |\det J(x)|$.
(See Appendix \ref{sec:imag-JdagJ-nonzero} for the details.)
Thus, neglecting the finite $\varepsilon$ effects,
the modulus of the Jacobian is included
in sampling, which may have certain advantage, but the price
one has to pay
seems overwhelming.

In the case of integrating $\tau$, on the other hand,
one obtains $\sqrt{\det g(x)} = C(x)\, |\det J(x)|$
neglecting the finite $\varepsilon$ effects,
where $C(x)\ge 0$ needs to be calculated and
taken into account in reweighting \cite{Fukuma:2020fez}.
Moreover, if the fixed point of the flow equation that contributes
to the integral resides on the original contour,
the volume element
$\sqrt{\det g(x)}$
of the worldvolume
vanishes
at that point, which causes an ergodicity 
problem.\footnote{In principle,
one can
deform the original contour so that
the fixed point is circumvented.
This is difficult in practice, however,
since the fixed point is not known {\it a priori}.}
In fact, this problem occurs because the induced metric
$g_{\mu\nu}(x)$
on the worldvolume
becomes singular 
at that point.
Obviously,
our HMC algorithm based on the Hamilton dynamics on the
original contour is totally free from such problems.

Note also that
the HMC algorithms on the deformed manifold
in Refs.\cite{Fukuma:2019uot,Fukuma:2020fez}
assume ${\rm Im} (J^\dag J)=0$, which is violated
upon discretization of the flow equation
as we show in Appendix \ref{sec:imag-JdagJ-nonzero}.
There are actually two places in which this assumption is used.
One of them is, as already mentioned above,
in the statement that
the HMC algorithm on the deformed contour includes $|\det J|$
in sampling.
The other is in constructing the basis vectors of $\bbC^N$ orthogonal to the
tangent space, which are used in solving the constrained motion on
the deformed manifold.
Therefore, the HMC algorithms in Refs.\cite{Fukuma:2019uot,Fukuma:2020fez}
suffer from systematic errors
due to discretization of the flow equation.
In contrast, nowhere in our HMC algorithm on the original contour
have we used ${\rm Im} (J^\dag J)=0$.
Hence, our algorithm is free from systematic errors
whatsoever.


\section{Practical applications}
\label{sec:examples}

In this section, we
consider
an important improvement of our algorithm
suggested also from our discussion in Section \ref{sec:relationship}.
This makes the discretization of the Hamilton equation
slightly more nontrivial as we describe below.
We test our algorithm by applying
it to
the time evolution of the wave function in the path integral
formalism.

\subsection{Introducing a mass parameter in the HMC}
\label{sec:mass-perameter}

As we have seen in Section \ref{sec:relationship},
the Hamilton dynamics on the deformed manifold
is equivalent to introducing a nontrivial kernel
$K(x)=\frac{1}{2} g^{-1}(x)$ in \eqref{H-equiv}
in the Hamilton dynamics on the original 
contour\footnote{Note, however, that
it is not straightforward to solve \eqref{Ham-eq-equiv}
numerically due to the term $\frac{\del K_{\nu\lambda}(x)}{\del x_\mu}$.}.
In a similar spirit, 
for practical applications, we find it very important
to generalize our HMC algorithm by introducing 
a $\tau$-dependent mass parameter $m(\tau)$ as
\begin{align}
  \tilde{Z}_W
  \label{Z-HMC-W-mtau}
  &=\int d\tau \, dp_\tau \, dx  \,  dp  \, e^{-H}  \ , \\
  H &= \frac{1}{2 m(\tau)} (p_i)^2 + \frac{1}{2} (p_\tau)^2
  + {\rm Re} S(z(x,\tau)) + W(\tau) \ .
\label{def_Ham-W-mtau}
\end{align}
Note first that integration over the momentum gives a factor
$m(\tau)^{N/2}$ in the partition function.
In the case of fixed $\tau$, this is just a constant factor,
while in the case of integrating $\tau$, it can be absorbed
into the definition of $W(\tau)$ in \eqref{def_Z_W}.
This also suggests that $m(\tau)$ should be chosen to be
proportional to 
the typical value of $|\det J(x,\tau)|^{2/N}$
for various $x$
with fixed $\tau$.

The need for this improvement can be understood from the fact
that the flow equation
typically maps a small region on the original contour to 
an exponentially large region on the deformed contour at large $\tau$,
where the scale factor is given by $|\det J(x,\tau)|^{1/N}$.
Introducing the mass in the HMC changes
the effective stepsize 
in solving
the Hamilton equation on the original contour 
by the factor of $1/\sqrt{m(\tau)}$.
Therefore, the above choice of $m(\tau)$ 
enables a random walk on the deformed manifold with 
almost uniform discretization.

From the Hamiltonian \eqref{def_Ham-W-mtau},
the Hamilton equation
is obtained as
\begin{align}
  \label{Ham-eq-mtau-x}
  \frac{dx_i(s)}{ds} &=
  \frac{1}{m(\tau)} \, p_i(s) \ , \\
  \frac{dp_i(s)}{ds} &=
  F_i(s) \ ,
  \label{Ham-eq-mtau} \\
    \label{Ham-eq-mtau-tau}
  \frac{d\tau(s)}{ds} &=
  p_\tau(s) \ , \\
  \frac{dp_\tau(s)}{ds} &=
   -   \left. \frac{d W(\tau)}{d \tau} \right|_{\tau = \tau(s)} 
  + F_\tau(s) 
  + \frac{1}{2 m(\tau)^2} \frac{d m(\tau)}{d\tau } (p_i)^2
\ ,
\label{Ham-eq-tau-mtau}
\end{align}
where the forces $F_i(s)$ and $F_\tau(s)$
are defined by \eqref{def-F} and \eqref{def-F-tau0}, respectively,
as before.

\subsection{Discretizing the Hamilton equation in the HMC}
\label{sec:discretizing-HMC}

In actual calculations, the Hamilton equation in the HMC
algorithm has to be discretized.
In the case of fixed $\tau$, the Hamiltonian
takes the canonical form $H=\frac{1}{2} p^2 + V(x)$
so that one can use the standard leap-frog discretization.
This respects the reversibility and the preservation
of the phase space volume \cite{Duane:1987de},
which guarantees the detailed balance.
In the case of integrating $\tau$, one should note
that the first term of
the Hamiltonian \eqref{def_Ham-W-mtau}
mixes
the momentum variables $p_i$ with $\tau$,
which
is now treated as one of the coordinate variables.
For this reason, we have to generalize the leap-frog discretization slightly.

Let us note first that the Hamilton equation takes the form
\begin{align}
  \frac{dp_\tau}{ds} = A(x,\tau,p) \ ,  \quad
  \frac{dx}{ds} = B(\tau, p)  \ , \quad
  \frac{d\tau}{ds} = C(p_\tau) \ , \quad
  \frac{dp}{ds} = D(x,\tau)  \ .
  \label{Ham-eq-formal}
\end{align}
We discretize this equation as
\begin{align}
  \label{Ham-eq-formal-discretized1}
  p_\tau (s_{n+1/2}) &= p_\tau (s_n) + \frac{\Delta s}{2} \,
  A(x(s_n),\tau(s_n),p(s_n))
  \ , \\
  x(s_{n+1/2}) &= x (s_n) + \frac{\Delta s}{2} \,  B(\tau(s_n),p(s_n))
  \ , \\
  \tau (s_{n+1/2}) &= \tau (s_n) + \frac{\Delta s}{2} \,  C(p_\tau (s_{n+1/2}))
\label{tau-evolve}
  \ , \\
  p(s_{n+1}) &= p(s_n) + \Delta s \, D(x(s_{n+1/2}),\tau(s_{n+1/2}))
  \ , \\
  \tau (s_{n+1}) &= \tau (s_{n+1/2}) + \frac{\Delta s}{2} \,  C(p_\tau (s_{n+1/2}))
  \ , \\
  x(s_{n+1}) &= x (s_{n+1/2}) + \frac{\Delta s}{2} \,  B(\tau(s_{n+1}),p(s_{n+1}))
  \ , \\
  p_\tau (s_{n+1}) &= p_\tau (s_{n+1/2}) + \frac{\Delta s}{2} \,
  A(x(s_{n+1}),\tau(s_{n+1}),p(s_{n+1}))
  \ ,
  \label{Ham-eq-formal-discretized4}
\end{align}
where $s_\nu = \nu \Delta s$ with $\nu$ being an integer or a half integer.
We repeat the above procedure for $n=0,1, \cdots , (N_s-1)$,
where $s_{\rm f} = N_s \Delta s$ represents the total fictitious
time for the Hamilton evolution.
One can prove
the reversibility and the preservation of the phase space volume
as in the standard leap-frog discretization.
Note that the procedure \eqref{Ham-eq-formal-discretized4} for $n=k$
and the procedure \eqref{Ham-eq-formal-discretized1} for $n=k+1$
can be combined into one step as
\begin{align}
  \label{Ham-eq-formal-discretized-combine}
  p_\tau (s_{k+3/2}) &= p_\tau (s_{k+1/2}) + \Delta s \,
  A(x(s_{k+1}),\tau(s_{k+1}),p(s_{k+1}))  
\end{align}
for $k=0 , 1 , \cdots , (N_s-2)$.

In actual simulation, it is useful to specify a finite range of $\tau$
over which we integrate in \eqref{def_Z_W}.  
If we realize this by 
choosing $W(\tau)$ in \eqref{def_Z_W} to be large outside
the range of $\tau$, the force in the Hamilton dynamics becomes very
large when $\tau$ gets out of the range.
Instead, we introduce walls at both ends of the range.
At the discretized level,
when $\tau$ gets out of the range
at \eqref{tau-evolve},
we go back to \eqref{Ham-eq-formal-discretized1},
flip the sign of $p_\tau (s_{n})$ there and continue.
One can easily prove that 
introducing this procedure
does not violate the reversibility of the Hamilton dynamics.

\subsection{Time evolution of the wave function by the path integral}
\label{sec:time-evolve-wf}

In this section, we test our algorithm by applying
it to
calculations in quantum mechanics
using the path integral
formalism (See, for instance, 
Refs.~\cite{Alexandru:2016gsd,Alexandru:2017lqr,Mou:2019tck} for
calculations of the real-time correlator.).
The fundamental object in this formalism is the transition amplitude
\begin{align}
{\cal A}(x_{\rm i}, t_{\rm i}; x_{\rm f}, t_{\rm f}) 
= \int {\cal D} x(t)\,  e^{i S[x(t)]} \ ,
  \label{QM-example}
\end{align}
where the integral is taken over 
all the paths $x(t)$ ($t_{\rm i} \le t \le t_{\rm f}$)
with the constraints $x(t_{\rm i})=x_{\rm i}$ and $x(t_{\rm f})=x_{\rm f}$.
The time evolution of the wave function is given by
\begin{align}
\Psi(x_{\rm f}, t_{\rm f}) 
&=
\int dx \, 
{\cal A}( x_{\rm f}, t_{\rm f}; x , t_{\rm i}) 
\Psi(x , t_{\rm i})
\nn \\
&=
\int {\cal D} x(t)\,  \Psi(x(t_{\rm i}) , t_{\rm i}) \,  e^{i S[x(t)]} \ , 
  \label{time-evolve-wf}
\end{align}
where the path integral is taken now
with the constraint $x(t_{\rm f})=x_{\rm f}$ only.

Here
we calculate
the time-evolved wave function
\eqref{time-evolve-wf} numerically
for the action
\begin{align}
S[x(t)] = \int dt \left\{ 
\frac{1}{2} \, m \left(\frac{dx}{dt}\right)^2 - V(x)  
\right\} \ ,
  \label{QM-action}
\end{align}
where we use $m=1$ and the quartic potential $V(x)=\frac{1}{4 !} x^4$.
We assume, for simplicity, a Gaussian form for the initial wave function 
\begin{align}
  \Psi(x , t_{\rm i}) &=
  \exp \left\{ - \frac{1}{4} \, \gamma \,  (x-\alpha)^2 \right\} \ ,
  \label{initial-wf}
\end{align}
where we set $\gamma=1$ and $\alpha=1$.

Let us first discretize the time $t$ 
as $t_n= (n-1) \, \epsilon$ 
and define
$x_n = x(t_n)$, where $n=1 , \cdots , (N+1)$.
We also define
$t_{\rm i} = t_1$, $x_{\rm i} = x_1$
and 
$t_{\rm f} = t_{N+1}$, $x_{\rm f} = x_{N+1}$.
Thus the time-evolved wave function $\Psi (x_{\rm f},t_{\rm f})$
can be obtained by evaluating the partition function
\begin{align}
  Z(x_{\rm f}) = \int
  dx \, e^{-S(x;x_{\rm f})} 
\label{Z-qm}
\end{align}
with the dynamical variables $x_i$ ($i=1,2, \cdots , N$),
where the action $S(x;x_{\rm f})$ is given by
\begin{align}
S(x;x_{\rm f}) &= \sum_{n=1}^{N} f(x_n , x_{n+1})
+ \frac{1}{4} \, \gamma \, (x_1 - \alpha)^2  \ , 
\label{QM-example-discrete}
\\
  f(x,y)
& \equiv   - i \epsilon 
\left\{ \frac{1}{2} \, m \left(\frac{x-y}{\epsilon}\right)^2 
-  \frac{V(x) + V(y)}{2} \right\} \ .
  \label{f-def}
\end{align}


\begin{figure}[t]
	\centering
	\includegraphics[width=7.5cm]{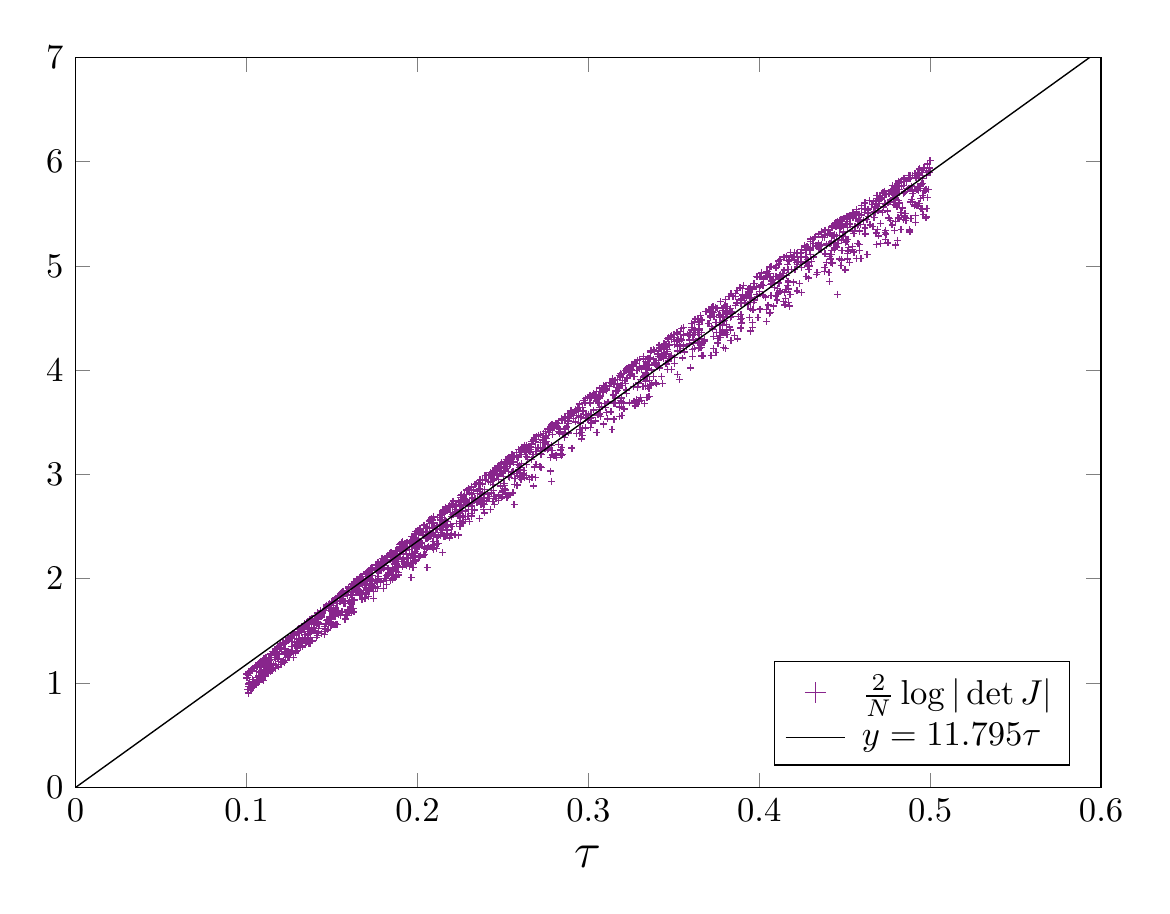}
	\includegraphics[width=7.5cm]{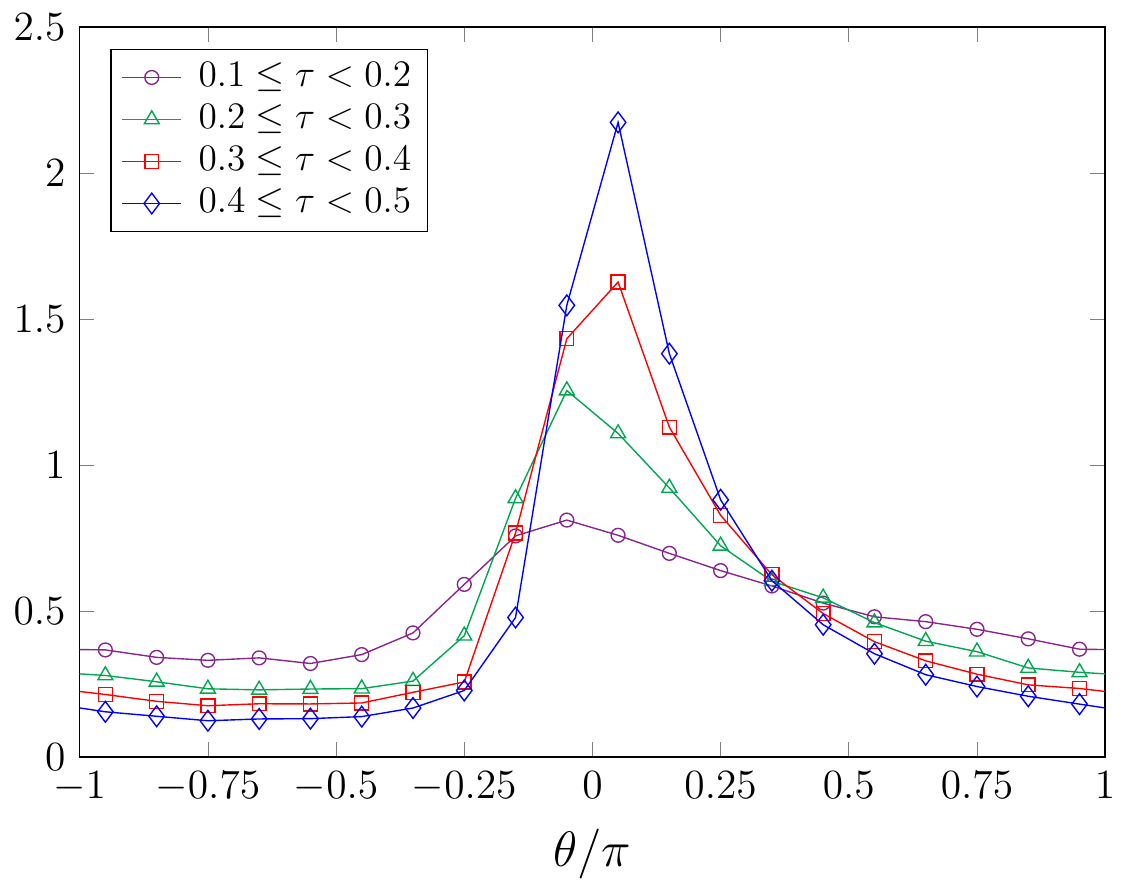}
	\caption{(Left) The quantity
           $\frac{2}{N} \log |\det J(z(x,\tau))|$
obtained for each configuration is plotted against $\tau$ for $N=9$.
The solid line represents a fit
to a linear function $c \, \tau$,
where $c=11.795$.
(Right) The phase distribution
of the reweighting factor in \eqref{gltm_ob-tau}
is plotted
for configurations with the value of $\tau$ in various regions for $N=9$.}
	\label{fig:mass_fit}
\end{figure}

\begin{figure}[t]
	\centering
	\includegraphics[width=9cm]{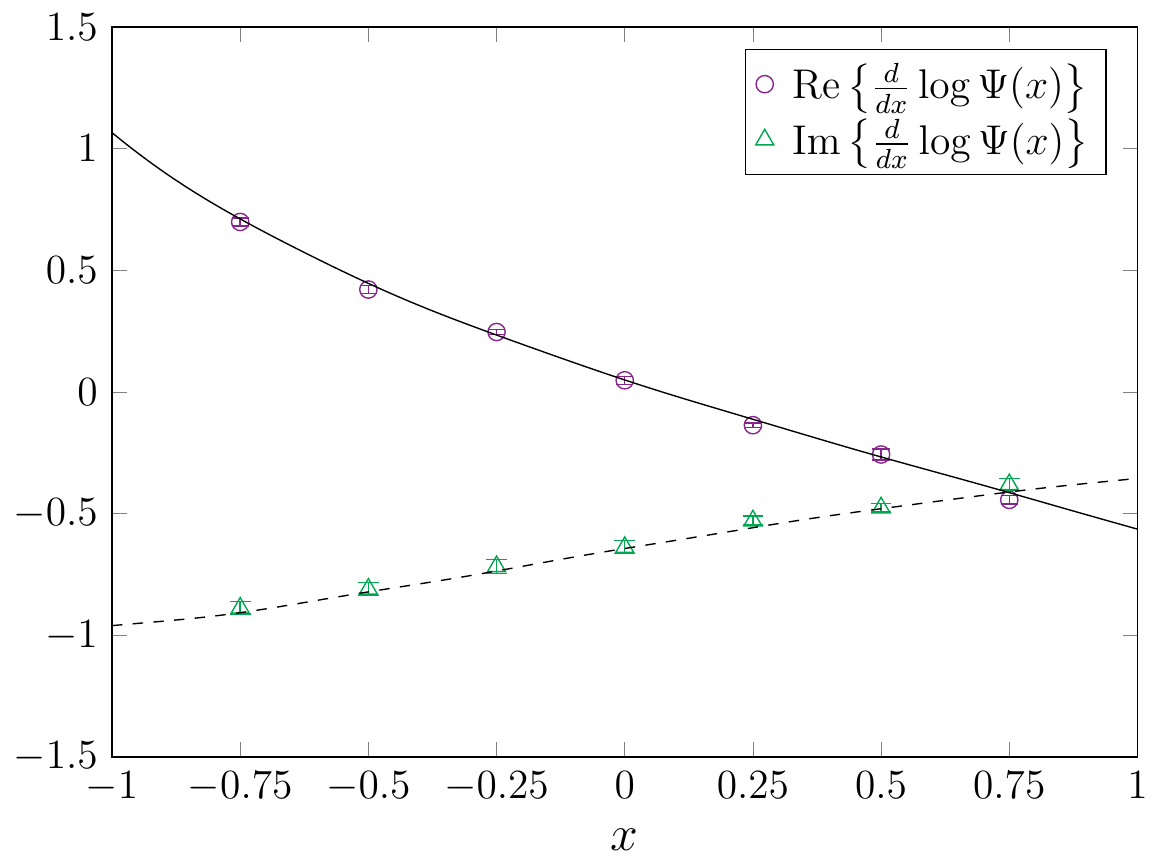}
	\caption{The real and imaginary parts
          of the log derivative of the wave function after time evolution with
          $t_{\rm f}=2$ is plotted for $N=9$.
          The solid line (real part) and
          the dashed line (imaginary part) represent 
          the results obtained 
          by diagonalizing the Hamiltonian, which corresponds to $N=\infty$.}
	\label{fig:quartic_wavefunction}
\end{figure}

We consider
the time evolution for $t_{\rm f}=2$, which is discretized by $N=9$.
The flow equation is discretized with $N_{\tau} = 10$ steps.
In the HMC algorithm, we use $N_s=15$ steps
with the total fictitious time 
$s_{\rm f} =0.25$.
We have confirmed that
the computational cost for
making one step in the HMC algorithm
scales linearly with $N$.

In Fig.~\ref{fig:mass_fit} (Left),
we plot $\frac{2}{N} \log |\det J(z(x,\tau))|$
obtained for each configuration against $\tau$.
We can fit the data points to a linear function $c \, \tau$
with $c=11.795$,
which is consistent with the fact
that the typical scale factor between
the original and deformed contours
grows
exponentially
with the flow time $\tau$.
According to our discussion in Section \ref{sec:mass-perameter},
we therefore use $m(\tau)=e^{c \, \tau }$ in the Hamiltonian \eqref{def_Ham-W-mtau}.
In reweighting \eqref{gltm_ob-tau},
we choose $\widetilde{W}(\tau)= \frac{1}{2}\,N\, c \, \tau$
in order to cancel the reweighting factor
$|\det J(x,\tau)|\sim m(\tau)^{N/2}$.
In Fig.~\ref{fig:mass_fit} (Right),
we plot the phase distribution
of the reweighting factor \eqref{gltm_ob-tau}
for configurations obtained within various regions of $\tau$.
We find that the peak of the distribution becomes sharper
as one goes to larger $\tau$,
which confirms that the sign problem becomes weaker as $\tau$ increases.

Instead of calculating 
the time-evolved wave function \eqref{time-evolve-wf} 
directly, we calculate its log derivative by
\begin{align}
\frac{\del}{\del x_{\rm f}} \log \Psi (x_{\rm f},t_{\rm f}) 
&=
- \left\langle 
\frac{\del}{\del x_{\rm f}} f(x_N, x_{\rm f})
\right\rangle_{x_{\rm f}} \ ,
  \label{time-evolve-wf-obs}
\end{align}
where
the expectation value is taken with respect to
the partition function \eqref{Z-qm}.
The results for $N=9$ are shown 
in Fig.~\ref{fig:quartic_wavefunction},
%
which is in good agreement
with the results obtained by diagonalizing the Hamiltonian
that corresponds to a continuous time evolution ($N=\infty$).


\section{Summary and discussions}
\label{sec:summary}

In this paper, we have proposed a new HMC algorithm 
for fast Lefschetz thimble calculations.
Unlike the existing HMC algorithms, which 
solve the Hamilton dynamics on the deformed contour,
we solve the Hamilton dynamics on the original contour
using the action evaluated at the point obtained by 
solving the holomorphic gradient flow equation.
The crucial point of our proposal is to
calculate the force in the Hamilton dynamics 
by backpropagating the force
on the deformed contour to that on the original contour,
analogously to the well-known idea that plays a central role 
in machine learning.
The computational cost in the present calculation is reduced
by a factor of the system size.

The possible ergodicity problem can be avoided by integrating
over the flow time \cite{Fukuma:2020fez},
which can be naturally implemented in our algorithm.
We discussed, in particular, that
the flow equation can be discretized even in this case without
causing systematic errors
unlike the existing HMC algorithms.
We have also discussed the relationship to the existing HMC algorithms
from the viewpoint of the Hamilton dynamics,
which revealed the existence of a nontrivial kernel in the 
kinetic term.
This kernel actually reproduces the volume element of the deformed manifold
upon integrating the auxiliary momentum variables.
Inspired by this observation, we introduced a flow-time dependent
mass parameter in the Hamilton dynamics, which turns out to be
important in practical applications.

Our algorithm is particularly useful in identifying the saddle points
and the associated thimbles that contribute to the integral in question.
This may shed light, for instance,
on quantum tunneling phenomena
from the viewpoint of
the real-time path 
integral \cite{Turok:2013dfa,Tanizaki:2014xba,Cherman:2014sba,Bramberger:2016yog,Mou:2019gyl}.
On the other hand, when one calculates the expectation values, 
one needs to calculate the Jacobian that appears in the reweighting factor.
This is time-consuming, but it can be done off-line only when one measures
the observables.
If one uses a crude estimator of the Jacobian
whose cost is only proportional
to the system size \cite{Alexandru:2016lsn},
the computational cost becomes
comparable to ordinary Monte Carlo
simulation for positive semi-definite weights.
We hope to apply our algorithm
to a large system in that way, although it would be certainly nice
if there is a better way to treat the Jacobian.

Last but not the least, the ``original contour'' in our algorithm
does not have to be the real axis but can be deformed in the
complex plane as far as one does not pass through singularities.
This points to the possibility
of combining the GTM with
various proposals for path
optimization \cite{Mori:2017pne,Mori:2017nwj,Alexandru:2018fqp,Bursa:2018ykf,Kashiwa:2018vxr,Alexandru:2018ddf,Detmold:2020ncp,Detmold:2021ulb},
which may be useful in reducing
the flow time required to solve the sign problem,
and hence in reducing the
effects
of the Jacobian in reweighting.

\subsection*{Acknowledgements}

We would like to thank Yuhma Asano,
Masafumi Fukuma, Yuta Ito,
Akira Matsumoto, Nobuyuki Matsumoto and
Neill C.\ Warrington
for valuable discussions.
The computations were carried out on
the PC clusters in KEK Computing Research Center
and KEK Theory Center. K.~S.\ is supported
by the Grant-in-Aid for JSPS Research Fellow, No.\ 20J00079.

\appendix

\section{Comments on the reduction of computational cost}
\label{sec:bp-jacobian}

In this appendix, 
we
provide a simple understanding
for the reduction of computational cost by backpropagation.

Let us first rewrite the differential equation \eqref{hgeJ} as
\begin{align}
  \frac{\partial}{\partial \sigma}
\begin{pmatrix}
    J(x,\sigma)  \\
   \bar{J} (x,\sigma)  
\end{pmatrix}
&=
{\cal H}(\sigma)
\begin{pmatrix}
   J(x,\sigma)    \\
    \bar{J}(x,\sigma)   
\end{pmatrix} \ ,
\label{hgeJ-matrix}
\end{align}
where we have defined the $2N \times 2N$ matrix ${\cal H}(\sigma)$ as
\begin{align}
{\cal H}(\sigma) &= 
  \begin{pmatrix}
    &
    \overline{H(z(x,\sigma))}  \\
        H(z(x,\sigma))  &   
  \end{pmatrix} \ .
\label{def-cal-H}
\end{align}
The solution to
the differential equation \eqref{hgeJ-matrix}
can be written down formally as
\begin{align}
\begin{pmatrix}
 J(x,\tau)  \\
 \bar{J} (x,\tau)  
\end{pmatrix}
&=
{\cal P} \exp \left( \int_0 ^\tau  d \sigma \, {\cal H}(\sigma) \right)
\begin{pmatrix}
   {\bf 1}_N \\
   {\bf 1}_N
\end{pmatrix} \ ,
\label{hgeJ-sol}
\end{align}
where ${\bf 1}_N$
is
the $N\times N$ unit matrix,
and ${\cal P} \exp$ represents the path-ordered exponential,
which ensures that ${\cal H}(\sigma)$ with smaller $\sigma$
comes on the right after Taylor expansion.
On the other hand, $F_j$ defined by \eqref{def-F2}
can be written as
\begin{align}
  F^{\top} &=
  \begin{pmatrix}
    f^{\top}  &   \bar{f}\, {}^{\top}
\end{pmatrix}
  \begin{pmatrix}
    J(x,\tau)  \\
    \bar{J} (x,\tau)  
  \end{pmatrix} \ .
\label{force-sol}
\end{align}
Plugging \eqref{hgeJ-sol} in \eqref{force-sol},
we obtain
\begin{align}
  F^{\top}
  &=
    \begin{pmatrix}
 f^{\top}  &   \bar{f}\, {}^{\top}
\end{pmatrix}
{\cal P} \exp \left( \int_0^\tau  d \sigma \, {\cal H}(\sigma) \right)
\begin{pmatrix}
   {\bf 1}_N \\
   {\bf 1}_N
\end{pmatrix} \ .
\label{force-sol2}
\end{align}
A naive method would be to obtain the Jacobian 
by \eqref{hgeJ-sol} and to 
obtain $F_j$ by \eqref{force-sol},
which corresponds to
taking the products in \eqref{force-sol2} from the right.

Let us consider how the backpropagation
calculates the same quantity \eqref{force-sol2}.
First, the ``force'' $f_i(\sigma)$ at flow time $\sigma$ defined by
\eqref{def-f-sigma} can be written as
\begin{align}
  \begin{pmatrix}
     f^{\top}(\sigma)  &
     \bar{f}{}^{\, \top}(\sigma)
\end{pmatrix}
  &=
    \begin{pmatrix}
    f^{\top}  &  \bar{f} {}^{\, \top}
\end{pmatrix}
        {\cal P} \exp \left( \int_{\sigma} ^\tau
        d \tilde{\sigma} \, {\cal H}(\tilde{\sigma}) \right) \ .
\label{force-backprop}        
\end{align}
This quantity satisfies the differential equation
\begin{align}
  \frac{d}{d \sigma}
  \begin{pmatrix}
    f^{\top}(\sigma)
    &  \bar{f}{}^{\, \top}(\sigma)
\end{pmatrix}
  &=  -  \begin{pmatrix}
    f^{\top}(\sigma)
    &  \bar{f}{}^{\, \top}(\sigma)
    \end{pmatrix}
 {\cal H}(\sigma) \ ,
\label{force-backprop-flow}        
\end{align}
which is equivalent to \eqref{force-backprop-flow-complex}.
By solving this backwards in $\sigma$
from $\sigma=\tau$ to $\sigma=0$
with the initial condition $f_j (\tau) = f_j$,
we obtain
\begin{align}
  \begin{pmatrix}
     f^{\top}(0)  &
     \bar{f}{}^{\, \top}(0)
\end{pmatrix}
  &=
    \begin{pmatrix}
    f^{\top}  &  \bar{f} {}^{\, \top}
\end{pmatrix}
        {\cal P} \exp \left( \int_{0} ^\tau
        d \tilde{\sigma} \, {\cal H}(\tilde{\sigma}) \right) \ .
\label{force-backprop-zero}        
\end{align}
Using this in \eqref{force-sol2}, we obtain
\begin{align}
  F^\top
  &=
    \begin{pmatrix}
     f^{\top}(0)  &
     \bar{f}{}^{\, \top}(0)
\end{pmatrix}
\begin{pmatrix}
   {\bf 1}_N \\
   {\bf 1}_N
\end{pmatrix}
=  f^{\top}(0) + \bar{f}{}^{\, \top}(0) 
\ .
\label{force-sol3}
\end{align}
Thus, the backpropagation amounts to reversing the order of
multiplications in \eqref{force-sol2},
which replaces the matrix-matrix products by the matrix-vector products,
thereby reducing the computational cost by a factor of O($N$).

The point just mentioned becomes clearer in the case of discretized
flow time discussed in Section \ref{sec:discretizing-flow-time}.
Corresponding to \eqref{hgeJ-matrix},
we can rewrite \eqref{hgeJ-discretized} in the form
\begin{align}
\begin{pmatrix}
    J(x,\sigma_{n+1})  \\
    \bar{J} (x,\sigma_{n+1} )
\end{pmatrix}
&=
\Big\{  1 + \varepsilon \, {\cal H}(\sigma_{n}) \Big\}
\begin{pmatrix}
   J(x,\sigma_n ) \\
    \bar{J} (x,\sigma_n)
\end{pmatrix}
\ ,
\label{hgeJ-matrix-discretized}
\end{align}
where the $2N \times 2N$ matrix ${\cal H}(\sigma)$ is defined
by \eqref{def-cal-H} as before.
The formal solution corresponding to \eqref{hgeJ-sol} can be written as 
\begin{align}
\begin{pmatrix}
    J(x,\tau)  \\
   \bar{J} (x,\tau)  
\end{pmatrix}
&= \prod_{k=1}^{N_{\tau}}
\Big\{  1 + \varepsilon \, {\cal H}(\sigma_{k-1}) \Big\}
\begin{pmatrix}
   {\bf 1}_N \\
   {\bf 1}_N
\end{pmatrix} \ ,
\label{hgeJ-sol-discretized}
\end{align}
where the product is taken in such a way that
a factor with smaller $k$ comes on the right.
Plugging \eqref{hgeJ-sol-discretized} in \eqref{force-sol},
we obtain
\begin{align}
  F^{\top}
  &=
    \begin{pmatrix}
 f^{\top}  &   \bar{f}\, {}^{\top}
\end{pmatrix}
\prod_{k=1}^{N_{\tau}}
\Big\{  1 + \varepsilon \, {\cal H}(\sigma_{k-1}) \Big\}
\begin{pmatrix}
  {\bf 1}_N \\
   {\bf 1}_N
\end{pmatrix} \ .
\label{force-sol2-discretized}
\end{align}

Corresponding to \eqref{force-backprop},
let us define
\begin{align}
  \begin{pmatrix}
     f^{\top}( \sigma_n)  &
     \bar{f}{}^{\, \top}( \sigma_n )
\end{pmatrix}
  &=
    \begin{pmatrix}
   f^{\top}  &  \bar{f}^{\, \top}
\end{pmatrix}
 \prod_{k=n+1}^{N_{\tau}}
\Big\{  1 + \varepsilon \, {\cal H}( \sigma_{k-1}) \Big\} \ ,
\label{force-backprop-discretized}        
\end{align}
which represents the force propagated backwards in $\sigma$.
Note that this quantity satisfies the
difference equation
\begin{align}
&  \begin{pmatrix}
    f^{\top}(\sigma_{n-1})
    &  \bar{f}{}^{\, \top}( \sigma_{n-1})
  \end{pmatrix}
  =
  \begin{pmatrix}
    f^{\top}(\sigma_{n})
    &  \bar{f} {}^{\, \top}(\sigma_{n})
\end{pmatrix}
  \Big\{  1 + \varepsilon \, {\cal H}(\sigma_{n-1}) \Big\} \ ,
\label{force-backprop-flow-discretized}        
\end{align}
which is equivalent to
\eqref{force-backprop-flow-complex0-discrete}.
By solving this
difference equation
with the initial condition $f_j (\tau) = f_j$,
we obtain the desired force as \eqref{force-final}.
Thus, in the discretized version,
the backpropagation simply amounts to
taking the products in \eqref{force-sol2-discretized}
from the left, which reduces the cost by a factor of O($N$)
compared with taking the products from the right.

\section{Some basic properties of the deformed manifold}
\label{sec:details}

In this appendix, we briefly review some basic properties of the deformed
manifold \cite{Fukuma:2020fez}, which is necessary in understanding
Section \ref{sec:relationship}.
We also 
provide a proof for the statement \eqref{project-N}.

The deformed manifold consists of points $z_i(x)$,
which can be obtained by solving either the continuous
or discretized version of the holomorphic gradient flow equation
for the flow time $\tau = x_0$
with the initial point given by $x_i$.
Therefore, the deformed manifold is
parametrized by $x_\mu$ ($\mu = 0, 1 , \cdots , N$),
and it is embedded in $\bbC^N$.
As we mentioned in Section \ref{sec:relationship},
our discussion applies to the case of fixed $\tau$ as well
by replacing the index $\mu = 0, 1 , \cdots , N$
with $i = 1 , \cdots , N$.

The tangent space at each point on the deformed manifold
is
a real linear space spanned by the basis vectors
\begin{align}
E_\mu^{\, i} &= \frac{\del z_i}{\del x_\mu} \ ,
  \label{def-E-vec}
\end{align}
and any element of it can be represented as
\begin{align}
  v_i &=
  c_\mu E_\mu^{\, i} \ , \quad \quad \mbox{where $c_\mu \in \bbR$} \ .
  \label{tangent-vector}
\end{align}
Since the deformed manifold is embedded in $\bbC^N$,
one can define the inner product of two tangent vectors
$u_i$ and $v_i$ as
\begin{align}
\langle u , v \rangle  &= {\rm Re} \left(\bar{u}_i v_i \right) \ .
  \label{inner-product}
\end{align}
Expanding the tangent vector $v_i$ as
\eqref{tangent-vector} and similarly for $u_i$ as
\begin{align}
  u_i &=
  b_\mu E_\mu^{\, i} \ ,
  \label{tangent-vector-u}
\end{align}
we can rewrite the inner product $\langle u , v \rangle$
in terms of the coefficients $b_\mu$ and $c_\mu$ as
\begin{align}
\langle u , v \rangle  &= g_{\mu\nu} b_\mu c_\nu \ ,
  \label{inner-product2}
\end{align}
where $g_{\mu\nu}$ is defined by
\begin{align}
g_{\mu \nu} &= {\rm Re} \left( \overline{E_\mu^{\, i}} \, E_\nu^{\, i}  \right) \ ,
  \label{def-g-tangent-vector}
\end{align}
which gives the induced metric \eqref{def-g} on the deformed manifold.

Let us consider a vector $w \in \bbC^N$ and project it onto the
tangent space. For that, let us we decompose $w$ as
\begin{align}
w_i &= a_\mu E_\mu ^{\, i} + y_i \ ,
  \label{project-w}
\end{align}
where $a_\mu \in \bbR$ is specified later,
and $y_i$ is a vector orthogonal to the tangent space; {\it i.e.},
\begin{align}
  {\rm Re} \left(\overline{E_\mu ^{\, i}} \, y_i \right) &= 0
  \quad\quad \mbox{for all $\mu$} \ .
\label{project-orthogonal}
\end{align}
From \eqref{project-w}, we obtain
\begin{align}
  {\rm Re} \left(\overline{E_\mu ^{\, i}} \, w_i \right)  &= a_\nu
  {\rm Re} \left(\overline{E_\mu ^{\, i}} \, E_\nu^{\, i} \right)  \\
  &= g_{\mu\nu} a_\nu \ .
  \label{project-w2}
\end{align}
Thus one obtains
\begin{align}
  a_\mu &= (g^{-1})_{\mu\nu}  {\rm Re}
  \left(\overline{E_\nu ^{\, i}} \, w_i \right) \ .
  \label{project-w3}
\end{align}
Plugging this in \eqref{project-w}, the
projection of $w$ onto the tangent space is given by
\begin{align}
  w_i \mapsto
  w_i  ' 
  &= E_\mu ^{\, i} 
  (g^{-1})_{\mu\nu}  {\rm Re} \left(\overline{E_\nu ^{\, j}} \, w_j \right) \ .
  \label{project-w4}
\end{align}
This confirms that \eqref{z-force-2}
is indeed the projection of the gradient force
$- \frac{\del V(z,\bar{z})}{\del \bar{z}_i} $
onto the tangent space.

Let us next prove \eqref{project-N}.
The left-hand side can be written as
\begin{align}
  {\rm Re} \left( \overline{\frac{\del z_i}{\del x_\alpha}} \, {\cal N}_i '
  \right)
  &= p_\lambda  {\cal M}_{\lambda \rho,\alpha} p_\rho \ ,
  \label{project-N-lhs}
\end{align}
where ${\cal M}_{\lambda \rho,\alpha}$ is defined by
\begin{align}
  {\cal M}_{\lambda \rho,\alpha}
  =&
  - \frac{1}{2} g_{\alpha\mu} K_{\mu\nu} \frac{\del K_{\lambda \rho}}{\del x_\nu} 
  + \frac{1}{2}
     K_{\nu \rho} g_{\alpha \mu} \frac{\del K_{\mu \lambda}}{\del x_\nu} 
  + \frac{1}{2}
  K_{\nu \lambda} g_{\alpha \mu} \frac{\del K_{\mu \rho}}{\del x_\nu}
  \nn \\
  &   + {\rm Re} \left(
     \overline{\frac{\del z_i}{\del x_\alpha}}
     \frac{\del^2 z_i}{\del x_\mu \del x_\nu} 
     \right)  K_{\nu\lambda} K_{\mu\rho} \ .
  \label{def-cal-M}
\end{align}
Let us recall here that $K=\frac{1}{2}\,g^{-1}$, which implies that
\begin{align}
  g_{\alpha\mu} K_{\mu\nu} \frac{\del K_{\lambda \rho}}{\del x_\nu} 
  &= \frac{1}{2} \frac{\del K_{\lambda \rho}}{\del x_\alpha}
  = - K_{\lambda \nu}
    \frac{\del g_{\mu\nu}}{\del x_\alpha}  K_{\mu \rho}  \ , 
  \\
  g_{\alpha \mu} \frac{\del K_{\mu \lambda}}{\del x_\nu} &=
  - \frac{\del g_{\alpha \mu}}{\del x_\nu}  K_{\mu \lambda} \ .
  \label{def-cal-M-2}
\end{align}
Using the definition \eqref{def-g} of the induced metric, we obtain
\begin{align}
  \frac{\del g_{\alpha \mu}}{\del x_\nu}
  &= 
   {\rm Re} \left(
  \frac{\del^2 z_i}{\del x_\nu \del x_\alpha}
  \overline{\frac{\del z_i}{\del x_\mu}}
  +   \frac{\del^2 z_i}{\del x_\mu \del x_\nu}
  \overline{\frac{\del z_i}{\del x_\alpha}} \right) \ .
  \label{del-g-x}
\end{align}
Thus \eqref{def-cal-M} becomes
\begin{align}
  {\cal M}_{\lambda \rho,\alpha}
  =&  K_{\nu\lambda} K_{\mu\rho}
\!
  \left\{
\frac{1}{2} 
   {\rm Re} \! \left(
  \frac{\del^2 z_i}{\del x_\mu \del x_\alpha}
  \overline{\frac{\del z_i}{\del x_\nu}}
  +   \frac{\del^2 z_i}{\del x_\nu \del x_\alpha}
  \overline{\frac{\del z_i}{\del x_\mu}} \right)
    \! -   \! \frac{1}{2} 
   {\rm Re} \! \left(
  \frac{\del^2 z_i}{\del x_\mu \del x_\alpha}
  \overline{\frac{\del z_i}{\del x_\nu}}
  +   \frac{\del^2 z_i}{\del x_\mu \del x_\nu}
  \overline{\frac{\del z_i}{\del x_\alpha}} \right) \right.
    \nn  \\
& \quad\quad \quad \quad  \left. - \frac{1}{2} 
   {\rm Re} \left(
  \frac{\del^2 z_i}{\del x_\nu \del x_\alpha}
  \overline{\frac{\del z_i}{\del x_\mu}}
  +   \frac{\del^2 z_i}{\del x_\mu \del x_\nu}
  \overline{\frac{\del z_i}{\del x_\alpha}} \right)
   + {\rm Re} \left(
  \frac{\del^2 z_i}{\del x_\mu \del x_\nu}
  \overline{\frac{\del z_i}{\del x_\alpha}} \right) \right\} = 0 \ ,
  \label{def-cal-M-3}
\end{align}
which completes the proof.

\section{Violation of ${\rm Im} (J^\dag J) = 0$ for the discretized flow}
\label{sec:imag-JdagJ-nonzero}

In this section, we point out that 
${\rm Im} (J^\dag J)=0$ holds only for the continuous flow
but not for the discretized one.
For the continuous flow equation, one obtains from \eqref{hgeJ},
\begin{align}
\frac{\partial}{\partial \sigma}(J^\dag J)_{ij}(x,\sigma)
&= \overline{J_{ki}(x,\sigma)} \, \overline{H_{kl}(z(x,\sigma))} \,
\overline{J_{lj}(x,\sigma)}
+ J_{ki}(x,\sigma) \, H_{kl}(z(x,\sigma)) \,
J_{lj}(x,\sigma)
\nn \\
&= 2 \, {\rm Re} \left\{ J_{ki}(x,\sigma) \, H_{kl}(z(x,\sigma)) \,
J_{lj}(x,\sigma) \right\} \ ,
\label{hgeJdag-J}
\end{align}
which implies that $(J^\dag J)_{ij}(x,\sigma) \in \bbR$
due to the initial condition $(J^\dag J)_{ij}(x,0)=\delta_{ij}$.

Note that in the first equality of \eqref{hgeJdag-J},
we need to use the Leibniz rule, which is violated
upon discretization of the flow equation.
Namely, we obtain 
\begin{align}
  & (J^\dag J)_{ij}(x,\sigma_{n+1})
  \nn \\
  =&
  \left\{
    \overline{J_{ki}(x,\sigma_n)} + \varepsilon \,
  J_{mi}(x,\sigma_n) \, H_{mk}(z(x,\sigma_n))
  \right\}
   \left\{
  J_{kj} (x,\sigma_{n_n}) + \varepsilon \,
  \overline{H_{kl}(z(x,\sigma_{n}))} \, \overline{J_{lj}(x,\sigma_{n})}
\right\}
\nn \\
=&  (J^\dag J)_{ij}(x,\sigma_n) + 
2 \, \varepsilon \, {\rm Re} \left\{
J_{mi}(x,\sigma_n) \, H_{mk}(z(x,\sigma_n)) \,  J_{kj} (x,\sigma_{n_n})  \right\}
\nn \\
& + \varepsilon^2 \, 
  J_{mi}(x,\sigma_n) \, H_{mk}(z(x,\sigma_n)) \, 
  \overline{H_{kl}(z(x,\sigma_{n}))} \, \overline{J_{lj}(x,\sigma_{n})} \ ,
\label{hgeJdag-J3}
\end{align}
where the last term on the right-hand side is not necessarily real.
Therefore, $(J^\dag J)_{ij}(x,\sigma_n)$ is not real at the discretized level.


\bibliographystyle{JHEP}
\bibliography{thimble-hmc}

\end{document}